\documentclass[twocolumn,aps,prl,showpacs,superscriptaddress]{revtex4-1}
\usepackage[latin9]{inputenc}
\usepackage{amsmath}
\usepackage[dvipdfmx]{graphicx}

\makeatletter

\usepackage{verbatim}
\usepackage{color}
\usepackage{subfigure}

\makeatother
\begin{document}

\title{Full Counting Statistics of Spin-Flip/Conserving Charge Transitions\\ in Pauli-Spin Blockade} 

\author{Sadashige Matsuo}
\email{sadashige.matsuo@riken.jp}
\affiliation{Department of Applied Physics, the University of Tokyo, 7-3-1 Hongo, Bunkyo-ku, Tokyo 113-8656, Japan}
\affiliation{JST, PRESTO, 4-1-8 Honcho, Kawaguchi, Saitama 332-0012, Japan}
\affiliation{Center for Emergent Matter Science, RIKEN, 2-1 Hirosawa, Wako-shi, Saitama 351-0198, Japan}

\author{Kazuyuki Kuroyama}
\affiliation{Department of Applied Physics, the University of Tokyo, 7-3-1 Hongo, Bunkyo-ku, Tokyo 113-8656, Japan}
\affiliation{Center for Emergent Matter Science, RIKEN, 2-1 Hirosawa, Wako-shi, Saitama 351-0198, Japan}

\author{Shunsuke Yabunaka}
\affiliation{Department of Physics, Kyushu University, Fukuoka 819-0395, Japan}

\author{Sascha R. Valentin}
\affiliation{Lehrstuhl f\"{u}r Angewandte Festk\"{o}rperphysik, Ruhr-Universit\"{a}t
Bochum, Universit\"{a}tsstra\ss e 150, D-44780 Bochum, Germany}

\author{Arne Ludwig}
\affiliation{Lehrstuhl f\"{u}r Angewandte Festk\"{o}rperphysik, Ruhr-Universit\"{a}t
Bochum, Universit\"{a}tsstra\ss e 150, D-44780 Bochum, Germany}

\author{Andreas D. Wieck}
\affiliation{Lehrstuhl f\"{u}r Angewandte Festk\"{o}rperphysik, Ruhr-Universit\"{a}t
Bochum, Universit\"{a}tsstra\ss e 150, D-44780 Bochum, Germany}

\author{Seigo Tarucha}
\email{tarucha@riken.jp}
\affiliation{Department of Applied Physics, the University of Tokyo, 7-3-1 Hongo, Bunkyo-ku, Tokyo 113-8656, Japan}
\affiliation{Center for Emergent Matter Science, RIKEN, 2-1 Hirosawa, Wako-shi, Saitama 351-0198, Japan}

\begin{abstract}
We investigate the full counting statistics (FCS) of spin-conserving and spin-flip charge transitions in Pauli-spin blockade regime of a GaAs double quantum dot. A theoretical model is proposed to evaluate all spin-conserving and spin-flip tunnel rates, and to demonstrate the fundamental relation between FCS and waiting time distribution. We observe the remarkable features of parity effect and a tail structure in the constructed FCS, which do not appear in the Poisson distribution, and are originated from spin degeneracy and coexistence of slow and fast transitions, respectively. This study is potentially useful for elucidating the spin-related and other complex transition dynamics in quantum systems.

\end{abstract}
\maketitle
The recent advances in charge sensing technologies using single electron transistors or quantum dots (QDs) have facilitated the tracking of charge dynamics, including charge tunneling, electron-phonon coupling, etc., with the resolution of single charge~\cite{fieldprl1993, lunature2003, schleserapl2004, vandersypenapl2004, reillyapl2007}.
Such charge dynamics can be used to reveal the microscopic mechanism of statistical or thermodynamical phenomena, such as the fluctuation theorem~\cite{sairaprl2012, kungprx2012, kungjap2013} and Maxwell demon engine~\cite{koskiprl2015, chidanatcommun2017}.
QDs have been extensively utilized as a tunable platform for investigating and controlling these phenomena.
Full counting statistics (FCS) is one of the most effective tools to analyze the charge dynamics, which yields the probability density $p(n,t)$ of $n$ transitions in a time window $t$.
FCS encodes all the cumulants, which include not only the mean but also the fluctuations and higher-order correlations ~\cite{levitovjmp1996, bagretsprb2003}.
Consequently, it has been used for investigating the cumulant asymmetry~\cite{gustavssonprl2006}, super-Poissonian properties~\cite{gustavssonprb2006}, and universal oscillation of the higher-order cumulants~\cite{flindtpnas2009} in a single QD, bidirectional counting and anti-bunching correlation in a double QD (DQD)~\cite{fujisawascience2006}, avalanche of the Andreev reflection events~\cite{maisiprl2014}, and optically detected single-electron tunneling~\cite{kurzmannprl2019}. However, experimental demonstration of FCS has been limited to QDs with few internal degrees of freedom.
In order to establish FCS for more complicated statistical phenomena, it is necessary to investigate QDs with more internal degrees of freedom, e.g., spin coupled quantum systems, or QDs exhibiting fast and slow transitions.
It may be noted that spin relaxation has been discussed in earlier papers~\cite{gustavssonprb2006, kurzmannprl2019}.
However, the dynamics of correlated spins in QDs has not been reported yet.

\begin{figure}[h]
\centering
\includegraphics[width=1.0\linewidth]{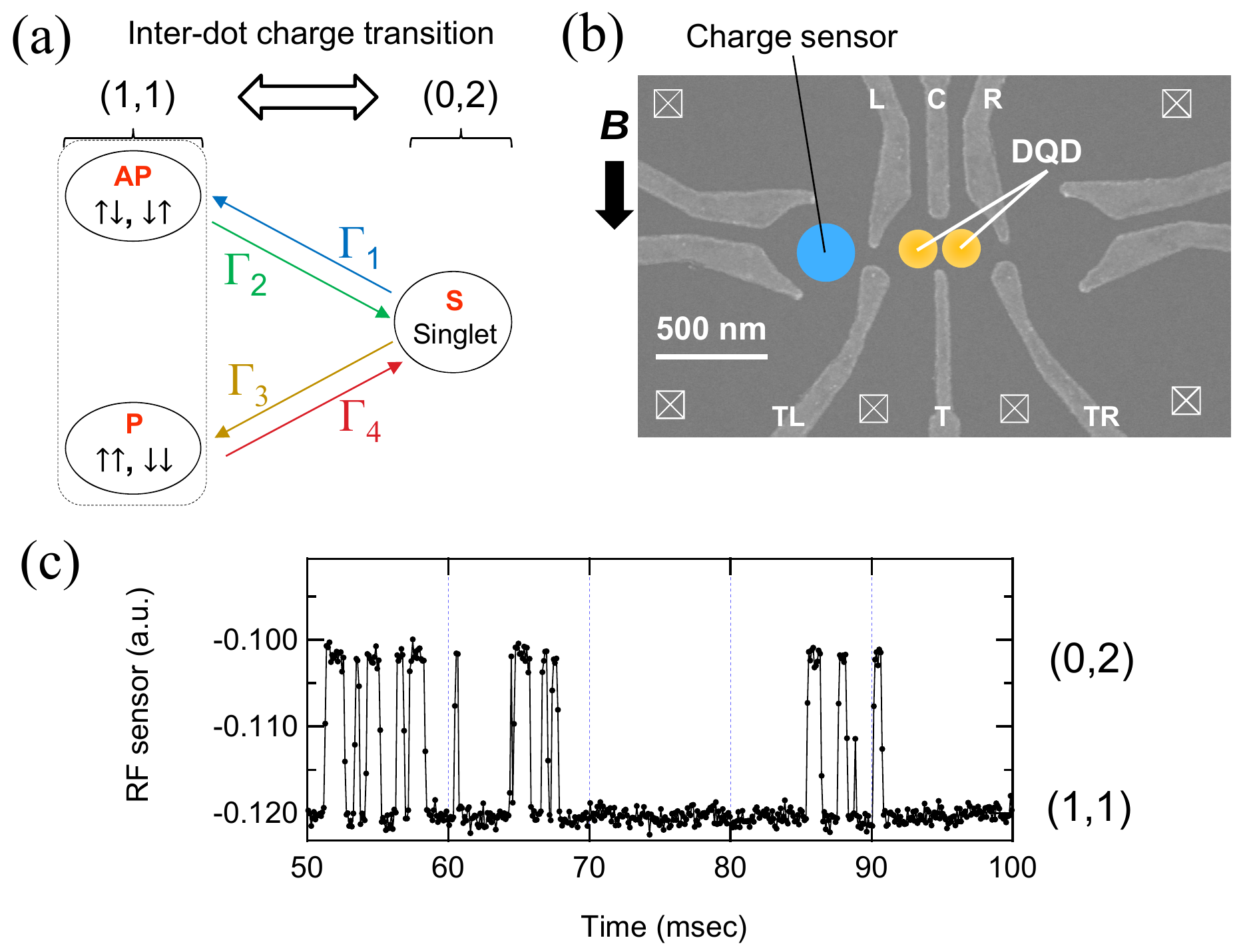}
\caption{
(a) Schematic diagram for tunneling events of a DQD in PSB. The possible spin configurations of (1,1) charge state are spin anti-parallel (AP) and spin parallel (P). 
The (0,2) charge state is spin-singlet (S). All the three possible states are connected by transitions with rates $\Gamma_1, \Gamma_2, \Gamma_3$, and $\Gamma_4$.
(b) Scanning electron microscopy (SEM) image of our DQD. The DQD and charge sensor QD are represented as yellow and blue circles, respectively.
(c) Typical time trace of $V_{rf}$. The jumps in $V_{rf}$ imply the inter-dot charge transitions between (1,1) and (0,2).
}
\end{figure}
In this work, we choose the Pauli-spin blockade (PSB) effect in a DQD~\cite{onoscience2002} to investigate the charge and spin dynamics because PSB is the simplest but most significant spin-correlated phenomenon that affects the electron dynamics in a DQD.
Real-time charge sensing of a DQD holding two electrons in PSB has been reported in earlier studies, which showed that the charge transitions can be classified into spin-flip and spin-conserving transitions~\cite{maisiprl2016, fujitaprl2016, hofmannprl2017}.
The spin-conserving transitions only occur when the two spins are anti-parallel, while the spin-flip transitions change the spin configuration.
Consequently, the spin configuration can be different even if the charge state is the same.
This additional degree of freedom complicates the charge dynamics. 

Here, we demonstrate the efficacy of FCS method in elucidating the microscopic dynamics of spin-conserving and spin-flip tunnels of a GaAs DQD holding two electrons in PSB. 
We construct the FCS experimentally and validate it theoretically using our model, which is used to derive all the necessary tunnel rates.
FCS is compared to the waiting time distribution (WTD), which has typically been utilized for evaluating the tunnel rate in the earlier studies.
The observed features in FCS of asymmetric tailing and parity effect, are then discussed. The proposed method and the results are potentially useful for understanding more complicated transition dynamics realized in multiple spin-correlated QDs.

For constructing the FCS, we experimentally obtained the real-time traces of charge transitions in the DQD in PSB. The DQD was made in a GaAs quantum well.
A scanning electron microscope (SEM) image of this DQD is shown in Fig. 1(b). Here, the target DQD is represented by yellow circles. We applied negative voltages on the gate electrodes indicated as L, C, R, TL, T, and TR, and tuned the DQD in resonance with the transition between (1,1) and (0,2) (see Supplemental Material (SM)).
Here, (0,2) indicates no electrons in the left QD and two electrons in the right QD.
Subsequently, we formed another QD (blue circle) as a charge sensor connected to the high-frequency resonance circuit.
We measured real-time traces of the rf sensor response $V_{rf}$ to probe the charge state. A typical real-time trace is shown in Fig. 1(c).
$V_{rf}$ exhibits almost binary values of $-0.10$ and $-0.12$, indicating the charge state of (0,2) and (1,1), respectively.
Therefore, the transitions between these two values indicate the inter-dot charge transitions.

The FCS of inter-dot charge transitions can be constructed from the acquired time traces. First, the raw traces are divided into many shorter time traces (time domains) with a span of $t$. Subsequently, the number of inter-dot transitions are counted in each time domain. For example, 5 time domains of $t=10$ ms duration can be created in Fig. 1(c). There are 10 transitions between 50 and 60 ms.
Finally, we estimate the probability density $p(n,t)$ from the number of time domains with $n$ transitions.

These constructed FCSs with $t=10$ and $50$ ms and $B=100$ mT are shown in Fig. 2.
Here, we find two remarkable features that are not observed in Poisson distribution, $(\Gamma t)^ne^{-\Gamma t}/n!$, which is represented by triangles with a single tunnel rate $\Gamma$ of 1.28 kHz (only for comparison).
First, the obtained FCS has a tail structure at lower $n$.
Second, a parity effect is evident about $n$; even $n$ exhibits higher probability than odd $n$.
To confirm that these two peculiar features originate from the electron dynamics and not from artifacts such as measurement noise, it is necessary to validate the experimental results with theoretical calculations.

To this end, we now introduce our theoretical model and apply it on the inter-dot transitions between (0,2) and (1,1) in PSB. The spin-conserving inter-dot charge transitions are allowed when the two electrons have opposite spins, but they are prohibited due to the Pauli exclusion principle when the two spins are parallel, and only the spin-flip transitions are allowed in this case.
Consequently, we classify (1,1) into anti-parallel (AP(1,1)) and parallel (P(1,1)) states of possible spin configurations. 
Now, high-energy excitations are absent, and we are only concerned with the bound state (0,2) whose spin configuration is spin-singlet (S(0,2)).
We define four tunnel rates as $\Gamma_1, \Gamma_2, \Gamma_3$, and $\Gamma_4$ between such possible states.
The transition diagram is schematically shown in Fig. 1(a), where
$\Gamma_1$ and $\Gamma_2$ are the spin-conserving tunnel rates, and $\Gamma_3$ and $\Gamma_4$ are the spin-flip rates.

We define $p_P(n,t)$, $p_{AP}(n,t)$, and $p_S(n,t)$ as the FCS of finding the final state as P(1,1), AP(1,1), and S(0,2) after $n$ transitions during the time span [$0,t$], respectively.
The momentum generation function is $P(\chi, t)=(\sum^\infty_{n=0}p_S(n,t)e^{in\chi},\sum^\infty_{n=0}p_{AP}(n,t)e^{in\chi},\sum^\infty_{n=0}p_P(n,t)e^{in\chi})^\tau$, where $\tau$ stands for transposition of a vector and $\chi$ represents the counting field~\cite{maisiprl2014}.
We assume that the transition follows a Markovian dynamics.
The time evolution equation of $P(\chi,t)$ can therefore be expressed as 
\begin{eqnarray}
\frac{dP(\chi, t)}{dt}&=&{\cal M}P(\chi,t)\nonumber\\
&=&\left(
\begin{array}{ccc}
-(\Gamma_1+\Gamma_3)&\Gamma_2e^{i\chi}&\Gamma_4e^{i\chi}\\
\Gamma_1e^{i\chi}&-\Gamma_2&0\\
\Gamma_3e^{i\chi}&0&-\Gamma_4
\end{array}
\right)P(\chi,t).
\end{eqnarray}
It may be noted that the experimental result in Fig. 2 corresponds to the case: $p(n,t)=p_S(n,t) + p_{AP}(n,t) + p_P(n,t)$.

\begin{figure}[t]
\centering
\includegraphics[width=1.0\linewidth]{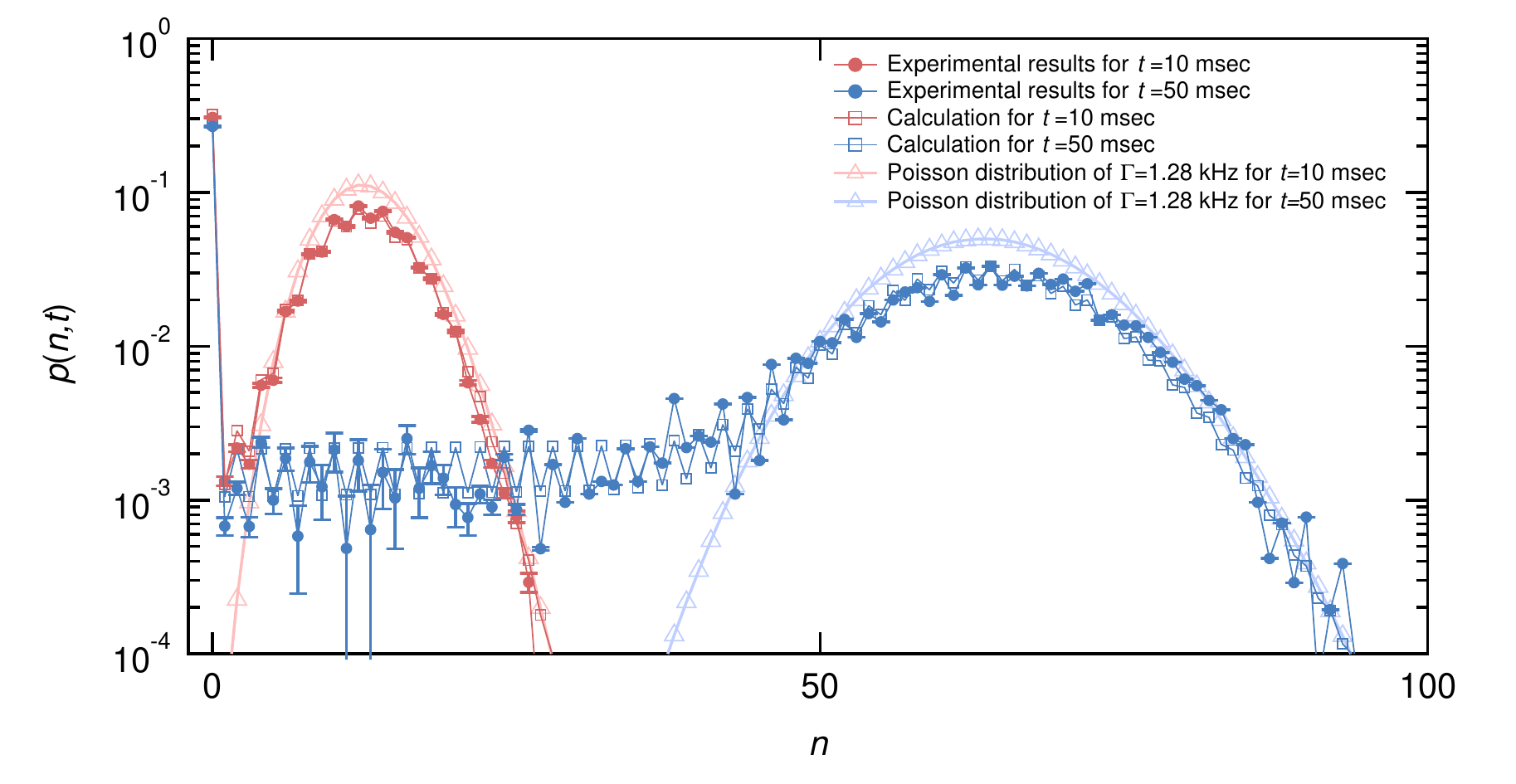}
\caption{
FCS in PSB. The red and blue circles (triangles) show the experimental (theoretical) results for $t=10$ and $50$ ms, respectively. The red and blue triangles indicate the Poisson distribution with $\Gamma =1.28$~kHz. 
}
\end{figure}
All the tunnel rates should be estimated to theoretically construct the FCS.
In the earlier studies, the WTD was used to evaluate the tunnel rates~\cite{maisiprl2016, fujitaprl2016, hofmannprl2017} but not $\Gamma_3$ because these studies focused on the exponents and not on the coefficients as discussed below. 
Furthermore, the waiting time in the blocked state P(1,1) is very long; therefore, a long data acquisition time is needed for the accurate estimation of $\Gamma_4$. 
We now focus on $p_S(0,t)$ and $p_{AP}(0,t)+p_P(0,t)$ because the charge state of either (0,2) or (1,1) can be detected.
The time evolution of probability distributions obeys Eq. (1) with $e^{i\chi}$ replaced by 0. Therefore, we obtain
\begin{eqnarray}
\left(
\begin{array}{c}
p_S(0,t)\\
p_{AP}(0,t)\\
p_P(0,t)
\end{array}
\right)=\left(
\begin{array}{c}
\frac{\Gamma_2\Gamma_4}{\Gamma_1\Gamma_4+\Gamma_2\Gamma_4+\Gamma_2\Gamma_3}e^{-(\Gamma_1+\Gamma_3)t}\\
\frac{\Gamma_1\Gamma_4}{\Gamma_1\Gamma_4+\Gamma_2\Gamma_4+\Gamma_2\Gamma_3}e^{-\Gamma_2t}\\ 
\frac{\Gamma_2\Gamma_3}{\Gamma_1\Gamma_4+\Gamma_2\Gamma_4+\Gamma_2\Gamma_3}e^{-\Gamma_4t}
\end{array}\right).
\end{eqnarray}
First, we can estimate $\Gamma_2$ and $\Gamma_4$ as the exponents in $p_{AP}(0,t)+p_P(0,t)$.
Subsequently, we can derive $\Gamma_1$ and $\Gamma_3$ from the coefficient ratio of the two exponential functions, $\Gamma_1\Gamma_4/\Gamma_2\Gamma_3$ in $p_{AP}(0,t)+p_P(0,t)$ and the exponent, $\Gamma_1+\Gamma_3$ in $p_S(0,t)$.
Consequently, we can estimate all the tunnel rates including $\Gamma_3$.

We now evaluate $p_S(0,t)$ and $p_{AP}(0,t)+p_P(0,t)$ from the time traces shown in Fig. 3(a).
Here, the solid lines represent the fitting results obtained by Eq. (2), which are in excellent agreement with the experimental results.
Consequently, all the tunnel rates can be determined as $(\Gamma_1,\Gamma_2,\Gamma_3,\Gamma_4)=(1.873$~kHz$,0.976$~kHz$,5.10$~Hz$,3.51$~Hz$)$.
It may be noted that $\Gamma_1/\Gamma_2 = 2$ due to the spin degeneracy ($\uparrow\downarrow$(1,1) and $\downarrow\uparrow$(1,1)) of AP(1,1) as previously reported~\cite{maisiprl2016, beckelepl2014}. 
We note that the spin-flip tunnels at $B=100$ mT are dominated by the spin-orbit interactions; therefore, we can ignore the intra-dot spin-flip tunnels due to the hyperfine interactions (see SM).

\begin{figure}[t]
\centering
\includegraphics[width=1.0\linewidth]{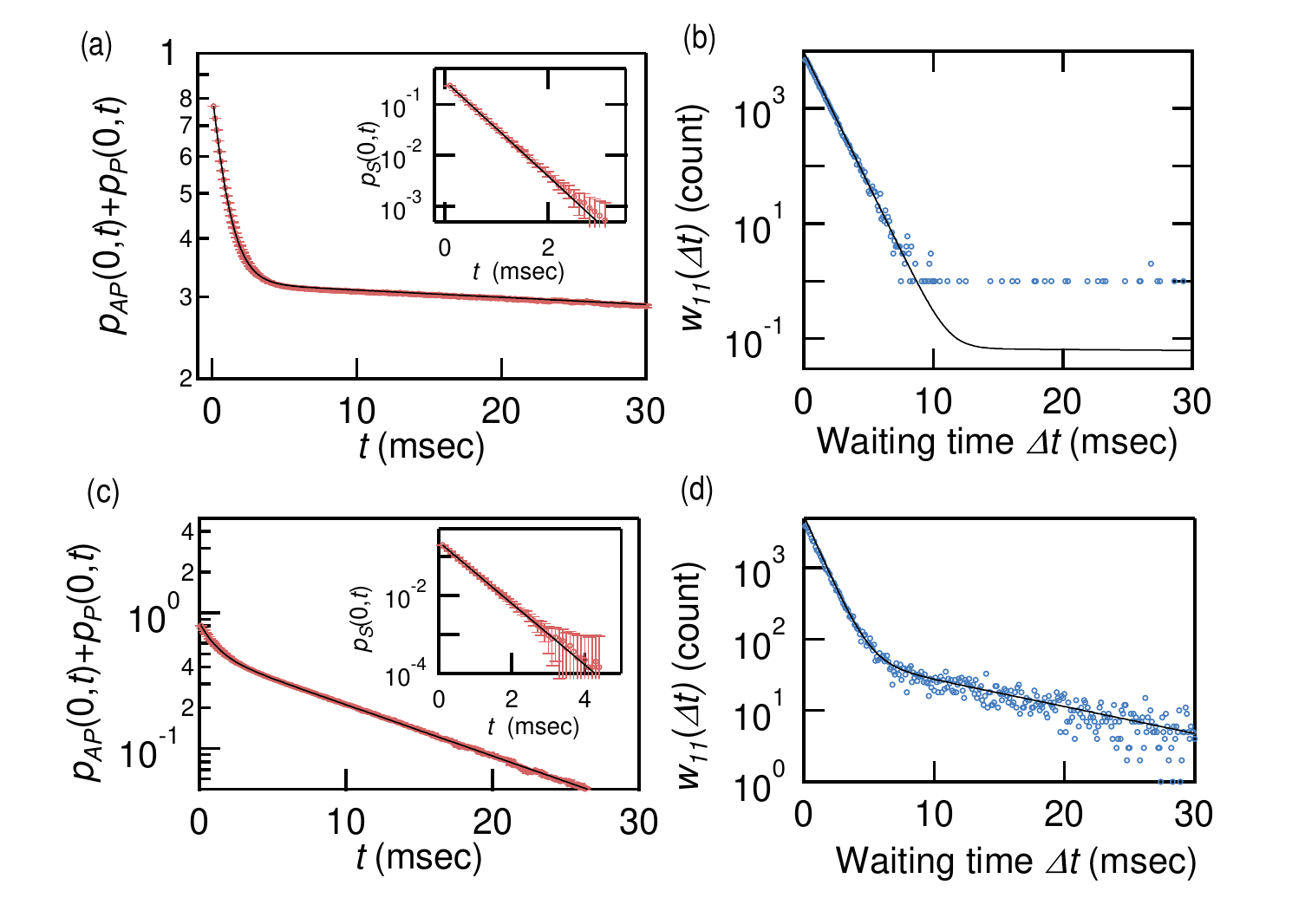}
\caption{Evaluation of tunnel rates.
(a) $p_{AP}(0,t)+p_P(0,t)$ is represented by the red circles. Inset shows $p_S(0,t)$. The numerical fitting results are denoted by the black curves.
(b) Blue circles represent $w_{11}(\Delta t)$ as a histogram of the waiting time $\Delta t$, which is evaluated from the real-time traces. The black curve indicates the numerically calculated result using the evaluated tunnel rates from $p(0,t)$.
(c) and (d) represent the same functions as (a) and (b) for different tunnel rates, respectively.
}
\end{figure}
From the estimated spin-flip rates, we can obtain $\Gamma_3/\Gamma_4 = 1.45$.
This ratio implies that there is an unintentional energy offset from the resonance condition.
This is because when these two tunnel rates are equal, the detailed balance condition implies $\Gamma_3/\Gamma_4 = 2\cosh (\Delta E_z/k_BT) \geq 2$, where $\Delta E_z,~k_B$, and $T$ are the Zeeman energy, Boltzmann constant, and temperature, respectively (see SM).

We now investigate the relation between $p(0,t)$ and WTD $w(\Delta t)$.
ThisWTD is the histogram of the waiting time $\Delta t$ in a certain charge state. 
Theoretically, the fundamental relation of $w(\Delta t)\propto d^2p(0,\Delta t)/d\Delta t^2$ is established~\cite{vyaspra1988, albertprl2012, haackprb2014} (see SM).
To demonstrate this relation, we focus on WTD for (1,1) charge state, $w_{11}(\Delta t)$ 
because both $p_S(0,t)$ and WTD for (0,2) are single exponential functions so that number of the differentiation is not explicitly demonstrated.
The relation for $w_{11}(\Delta t)$ is written by $w_{11}(\Delta t)\propto d^2(p_{AP}(0, \Delta t)+p_{P}(0, \Delta t))/d\Delta t^2$, resulting in $w_{11}(\Delta t) \propto \Gamma_1\Gamma_2e^{-\Gamma_2\Delta t}+\Gamma_3\Gamma_4e^{-\Gamma_4\Delta t}$.

The histogram of $\Delta t$ (proportional to $w_{11}(\Delta t)$) is shown as blue circles in Fig. 3(b).
The histogram exhibits unity or zero values for $\Delta t >10$ ms because the acquired time trace number is not large enough due to the slow spin-flip rates and short measurement time.
In this case, the evaluation of tunnel rate using $w_{11}(\Delta t)$ is not accurate compared to that using $p_{AP}(0,t)+p_P(0,t)$, which is confirmed by the theoretical results.
The ratio of coefficients for the two exponential functions in $w_{11}(\Delta t)$, i.e., $\Gamma_3\Gamma_4/\Gamma_1\Gamma_2 << 1$ is much smaller than the ratio $\Gamma_1\Gamma_4/\Gamma_2\Gamma_3 \approx 2$ in $p_{AP}(0,t)+p_P(0,t)$.
Therefore, the required measurement time to guarantee the evaluation accuracy is longer for WTD than for FCS with $n=0$.
The black line in Fig. 3(b) shows the calculated $w_{11}(\Delta t)$ from the tunnel rates, which cannot reproduce the experimental results.

We obtained the values of $p_{AP}(0,t)+p_P(0,t),~p_S(0,t)$, and $w_{11}(\Delta t)$ at different tunnel rates $(\Gamma_1,\Gamma_2,\Gamma_3,\Gamma_4) = (1.58$~kHz$,0.955$~kHz$,236$~Hz$,87.7$~Hz$)$, which are shown in Figs. 3(c) and (d). 
The theoretically calculated value of $Cd^2(p_{AP}(0, \Delta t)+p_{P}(0, \Delta t))/dt^2$, which is shown as the black line in Fig. 3(d), is in complete agreement with the experimentally obtained histogram. 
Here, the proportionality coefficient $C$ is a fitting parameter. 
Therefore, we have confirmed the fundamental relation between FCS with $n=0$ and WTD.
This demonstration implies that FCS with $n=0$ and the relation allow to reproduce the WTD without a long measurement time to accumulate the traces.

Finally, we calculate the FCS including $n(\neq 0)$ with the estimated tunnel rates based on Eq. (1), which yields $P(\chi ,t)=e^{{\cal M}t}P_0$.
$P_0$ is probability with the stationary condition, which is calculated from Eq. (1) with $dP(\chi,t)/dt=0$ and $\chi =0$.
This results in Eq. (2) with $t=0$.
The open squares in Fig. 2 are the calculation results using the estimated rates in Fig. 3(a) (see SM for details).
It is evident that the numerical simulations reproduce the experiments perfectly, including the lower $n$ tail structure and the parity effect.
This agreement validates that our model based on FCS can explain the transition dynamics of spin-flip and spin-conserving transitions in PSB. It further indicates that the tail structure and the parity effect in Fig. 2 are originated from the electron dynamics. Therefore, we have to establish these physical origins. First, we assign the lower $n$ tail to the slow spin-flip rates.
As indicated by Eq. (2), $p_S(0,t)$ and $p_{AP}(0,t)$ rapidly decay with $t$ as compared to $p_P(0,t)$. This implies that many spin-conserving transitions occur even in the small span $t$, while the spin-flip transitions occur rarely.
Here, the time domains that contain the spin-conserving transitions contribute to the peak at large $n$, and those containing the finite spin-flip transitions in addition to the spin-conserving transitions contribute to the long slope at smaller $n$. 
This is also supported by the FCS result at fast spin-flip rate because the corresponding probability of the tail structure becomes much larger than that at the slow spin-flip rate (see SM).

\begin{figure}[t]
\centering
\includegraphics[width=1.0\linewidth]{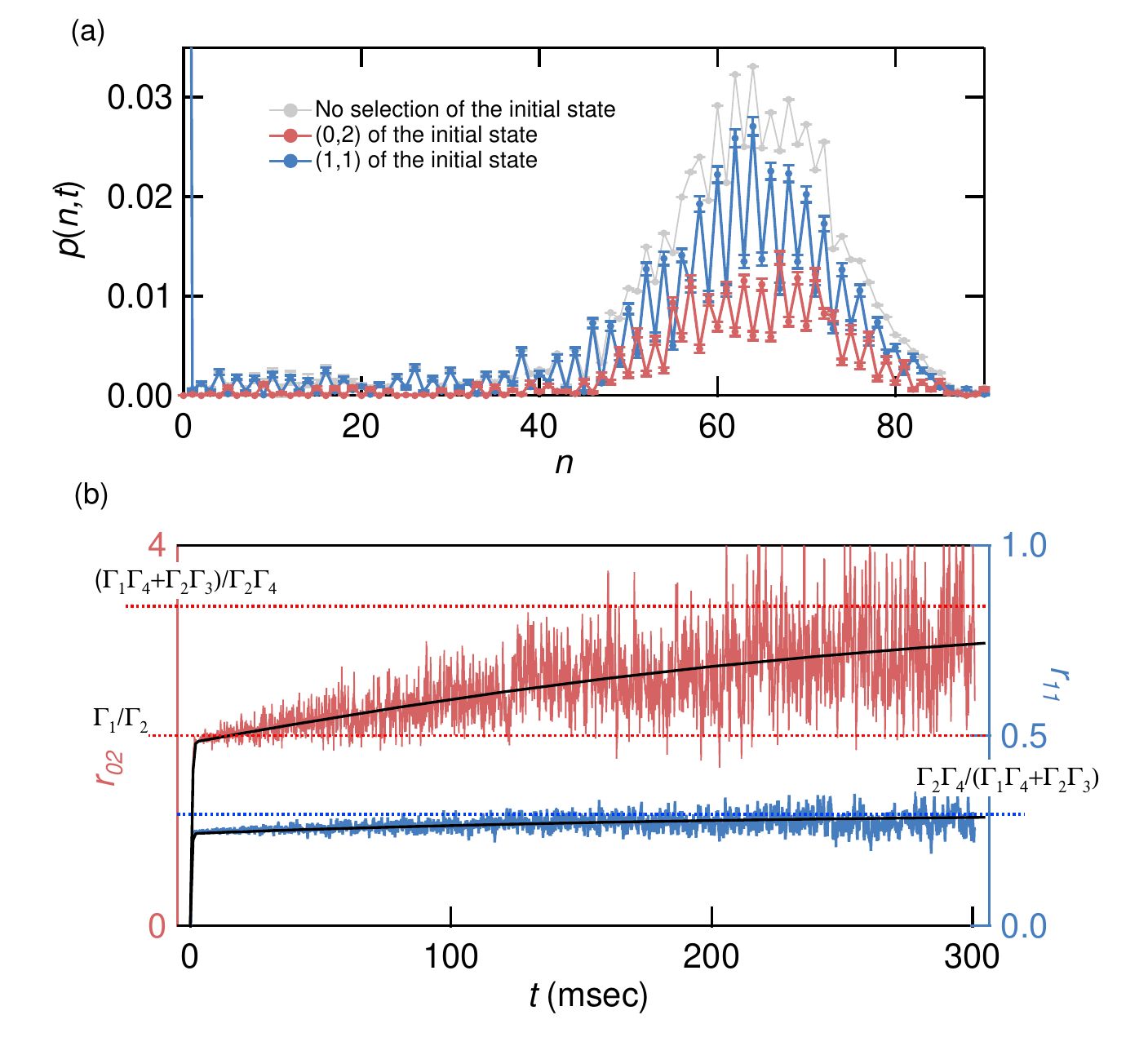}
\caption{
(a)FCS with and without the selection of the initial state. The gray circles indicate FCS without selection, while the blue and red circles indicate FCS with (0,2) and (1,1) as the initial state, respectively. The parity effect reverses for the different initial states.
(b)The red and blue lines represent the ratio of the odd $n$ probability to the even $n$ probability for the (0,2) and (1,1) initial states, respectively. The black lines show the calculated results based on our theoretical model.
}
\end{figure}
We reconstructed the FCS of the time domains with the same initial states to elucidate the origin of the parity effect.
The red and blue circles in Fig. 4(a) indicate the FCS constructed using the time domains with the initial state as (0,2) and (1,1) with $t=50$ ms, respectively. 
The grey circles are equivalent to the blue circles in Fig. 2.
It is evident here that the parity effect on the red circles is opposite to that on the blue ones. 
This can be understood in terms of the equilibration of the initial states.
The selected initial state, i.e., (0,2) or (1,1) is equilibrated into the (0,2) and (1,1) states after a long time with probabilities
$\frac{\Gamma_2\Gamma_4}{\Gamma_1\Gamma_4+\Gamma_2\Gamma_4+\Gamma_2\Gamma_3}\approx 1/5$ and 
$\frac{\Gamma_1\Gamma_4+\Gamma_2\Gamma_3}{\Gamma_1\Gamma_4+\Gamma_2\Gamma_4+\Gamma_2\Gamma_3}\approx 4/5$, respectively.
Then, the charge state tends to be (1,1) rather than (0,2) due to the higher spin degeneracy in (1,1).
Herein, the probability of odd $n$ becomes larger for the initial state (0,2) because the (0,2) state evolves to (1,1) after the odd $n$ transitions. 
On the contrary, the probability of even $n$becomes larger when the initial state is (1,1), resulting in an opposite parity effect to the case with (0,2) as the initial state.
The parity effect in FCS with no initial state selection is dominated by (1,1) initial state because the corresponding probability is larger than that for the (0,2) case, as seen in Fig. 3(a).

The time evolution of the parity effect can be explained in terms of $r_{02}$ and $r_{11}$, defined as
\begin{eqnarray}
\frac{\sum_{m=0}^\infty p_S(2m+1,t)}{\sum_{m=0}^\infty p_S(2m,t)}, {\rm and}\notag\\
\frac{\sum_{m=0}^\infty (p_{AP}(2m+1,t)+p_P(2m+1,t))}{\sum_{m=0}^\infty (p_{AP}(2m,t)+p_P(2m,t))},\notag
\end{eqnarray}
which are plotted as blue and red lines in Fig. 4(b), respectively.
The numerical calculations (black lines) are in excellent agreement with the experiments.
$r_{02}$ approaches $\Gamma _1/\Gamma_2$ around $t=1$ ms $\approx 1/\Gamma _2$, and then it becomes $(\Gamma_1\Gamma_4+\Gamma_2\Gamma_3)/\Gamma_2\Gamma_4$ around $t\approx 1/\Gamma _4$.
This is because the spin-conserving tunnels between (0,2) and AP(1,1) occur initially due to the larger rate.
Then the spin-flip tunnels generate the transitions between (0,2) and P(1,1) with the smaller rates. 
$r_{11}$ evolves as $\Gamma_2\Gamma_4/(\Gamma_1\Gamma_4+\Gamma_2\Gamma_3)$.
Such time evolution reflects the equilibration of the initial state, which finally saturates at the ratio corresponding to the equilibrium condition.

In conclusion, we analyzed the FCS of spin-conserving and spin-flip charge transitions in PSB both experimentally and theoretically.
The proposed model facilitated the estimation of all the necessary tunnel rates, which revealed that only one of the two spin-parallel states is significant for the spin-flip transitions in PSB. 
Then we demonstrated the fundamental relation between FCS and WTD, which means that WTD can be reproduced from FCS with $n=0$ even if a measurement time is short.
Further, we constructed the FCS and found two peculiar features: the tail structure and parity effect, which reflected the slow spin-flip tunnel rates and higher spin degeneracy in (1,1), respectively.
We believe that our results provides a powerful tool for understanding the transition dynamics of complex spin-correlated phenomena, which includes higher degeneracy, several tunnel rates, etc.

This work was partially supported by 
the Grant-in-Aid for Scientific Research (B) (grant number: JP18H01813), 
the Grant-in-Aid for Scientific Research (S) (grant numbers: JP26220710 and JP19H05610), JSPS Program for Leading Graduate Schools (ALPS),
JSPS Research Fellowship for Young Scientists (grant numbers: JP16J03037 and JP19J01737), 
the Grant-in-Aid for Scientific Research on Innovative Area Nano Spin Conversion Science (grant number: JP17H05177), JST CREST (grant number: JPMJCR15N2), and JST PRESTO (grant number: JPMJPR18L8).

S.M. and K.K. contributed equally to this work.

\bibliographystyle{apsrev4-1}
%

\end{document}